\documentclass[12pt]{article}
\usepackage{jheppub}
\usepackage{mathtools,amssymb,amsthm}
\usepackage[utf8]{inputenc}
\usepackage{enumitem}
\usepackage[dvipsnames]{xcolor}
\usepackage{graphicx}
\usepackage{multirow}
\usepackage[normalem]{ulem}
\usepackage{longtable}
\allowdisplaybreaks
\usepackage{slashed}
\usepackage{bbm}

\hypersetup{
    colorlinks = true,
    allcolors = {Maroon}
}

\newcommand{\be}{\begin{equation}}
\newcommand{\ee}{\end{equation}}
\def \bea {\begin{eqnarray}}
\def \eea {\end{eqnarray}}
\def\bal#1\ea{\begin{align}#1\end{align}}
\def\bad#1\ead{\begin{aligned}#1\end{aligned}}
\def\bg#1\eg{\begin{gather}#1\end{gather}}
\def\bm#1\em{\begin{multline}#1\end{multline}}
\def\bmd#1\emd{\begin{multlined}#1\end{multlined}}

\newcommand{\ignore}[1]{}

\def\d{\delta}

\def\i{\mathrm{i}}

\def \bal#1\eal  {\begin{align} #1 \end{align}}
\def \bga#1\ega  {\begin{gather} #1 \end{gather}}
\def\({\left(}
\def\){\right)}
\def\[{\left[}
\def\]{\right]}
\def\<{\left\langle}
\def\>{\right\rangle}
\def\d{\mathrm{d}}

\newcommand{\eim}{\end{itemize}}
\newcommand{\beq} {\begin{equation}}
\newcommand{\eeq} {\end{equation}}
\newcommand{\bc}{\begin{center}}
\newcommand{\ec}{\end{center}}

\newcommand{\nn} {\nonumber\\}

\newcommand{\pd} {\partial}

\newcommand{\ai}{{\alpha}}

\newcommand{\di}{{\delta}}
\newcommand{\ri}{{\rho}}
\newcommand{\si}{{\sigma}}
\newcommand{\li}{{\lambda}}
\newcommand{\ti}{{\tau}}

\newcommand{\oi}{\omega}
\newcommand{\epi}{\epsilon}

\newcommand{\Di}{\Delta}

\newcommand{\pr}{{\prime}}

\begin{document}

\title{Higher-spin localized shocks}
\author[a]{Diandian Wang,}
\author[b]{and Zi-Yue Wang}
\affiliation[a]{Center for the Fundamental Laws of Nature, Harvard University,\\
Cambridge, MA 02138, USA}
\affiliation[b]{Department of Physics, University of California,\\
Santa Barbara, CA 93106, USA}
\emailAdd{diandianwang@fas.harvard.edu}
\emailAdd{zi-yue@physics.ucsb.edu}

\abstract{In the context of AdS/CFT, gravitational shockwaves serve as a geometric manifestation of boundary quantum chaos. We study this connection in general diffeomorphism-invariant theories involving an arbitrary number of bosonic fields. Specifically, we demonstrate that theories containing spin-2 or higher-spin fields generally admit classical localized shockwave solutions on black hole backgrounds, whereas spin-0 and spin-1 theories do not. As in the gravitational case, these higher-spin shockwaves provide a means to compute the out-of-time-order correlator. Both the Lyapunov exponent and the butterfly velocity are found to universally agree with predictions from pole skipping. In particular, higher-spin fields lead to a Lyapunov exponent that violates the chaos bound and a butterfly velocity that may exceed the speed of light. 
}

\maketitle

\section{Introduction}

AdS/CFT \cite{Maldacena:1997re,Witten:1998qj} is a strong-weak duality. A practical value of this fact is that something complex and challenging to study on one side can sometimes be mapped to a rather simple counterpart on the other side. Holographic entanglement entropy is one such example. An intrinsically quantum mechanical boundary quantity is dual to an extremely simple classical object in the bulk, the area of an extremal surface known as the Ryu-Takayanagi (RT) surface \cite{Ryu:2006bv,Ryu:2006ef}. The (late-time) out-of-time-order correlator (OTOC) is another example. In quantum systems, it characterizes quantum chaos, and its bulk dual is a well-known object in classical gravity: the shockwave \cite{Shenker:2013pqa,Roberts:2014isa,Roberts:2016wdl,Shenker:2013yza,Shenker:2014cwa}. 

A gravitational shockwave is an exact solution in General Relativity with a distributional nature. Its metric differs from a smooth one by a Dirac delta function which has support on a codimension-one surface. The shockwave was historically constructed by studying the geometrical backreaction in response to highly energetic particles \cite{Aichelburg:1970dh, Dray:1984ha, Sfetsos:1994xa}. For shockwaves in pure AdS, the simplicity and analyticity of such solutions make them powerful tools for studying many properties of the gravitational theory and their dual CFTs \cite{Horowitz:1999gf,Cornalba:2006xk,Cornalba:2006xm,Cornalba:2007zb,Nozaki:2013wia,Camanho:2014apa,Asplund:2014coa,Hartman:2015lfa,Afkhami-Jeddi:2016ntf,Afkhami-Jeddi:2017rmx,Fitzpatrick:2019efk}.

We will be interested in those that travel along the black hole horizon, primarily because black holes are dual to thermal states of the boundary CFT \cite{Maldacena:2001kr}, and we are interested in studying the boundary system at finite temperatures. Such a shockwave describes the backreaction of a particle that has been falling into the black hole for an infinite amount of time and is therefore infinitely boosted by the black hole itself. The location of the particle breaks the symmetry in the transverse directions, making the shockwave \emph{localized} in this sense \cite{Roberts:2014isa}. As this work focuses exclusively on localized shockwaves, we will henceforth refer to them simply as shockwaves.

In the bottom-up approach to AdS/CFT, different bulk theories can be studied even though their precise duals are not known. The AdS/CFT dictionary empowers us to compute corresponding CFT quantities from bulk data. By analyzing assorted bottom-up models in the bulk, it is possible to learn about the rigidity of certain properties of holographic CFTs under various deformations. The simplest modifications to Einstein gravity include the addition of matter fields and higher-curvature corrections. 

For many gravitational phenomena, adding minimally coupled low-spin matter to Einstein gravity does not usually destroy the correspondence, though it could change the quantitative details. One such example is the RT surface mentioned earlier, where adding minimally coupled low-spin matter does not change the formula as long as the action does not contain too many derivatives. Perturbative higher-derivative corrections also do not destroy the existence of such a correspondence, though the RT functional itself does receive (perturbative) corrections \cite{Dong:2013qoa,upcomingRT}. It turns out that the shockwave/OTOC correspondence is similar, i.e., low-spin fields and higher-curvature corrections change the shockwave metric and therefore details of the OTOC, but the existence of such a correspondence is not destroyed \cite{Mezei:2016wfz,Alishahiha:2016cjk,Qaemmaqami:2017bdn,Li:2017nxh,Dong:2022ucb}.

In this work, we ask the following question: How much can we say about these dual descriptions beyond the ``good matter" comfort zone by involving isolated higher-spin fields in our bulk theory? By ``isolated'', we mean a finite number of such fields, in contrast to theories of higher-spin gravity which feature an infinite, correlated tower of higher-spin fields. Isolated higher-spin fields are usually perceived quite negatively. For one thing, they are known to violate causality \cite{Camanho:2014apa}. Massless ones are also known to be forbidden by symmetry \cite{Bekaert:2010hw,Maldacena:2011jn}. As a result, pursuing this direction might seem unappealing. Nevertheless, the holographic dictionary \cite{Gubser:1998bc,Witten:1998qj,Banks:1998dd} is still applicable even when the theories (on both sides) are pathological, at least in the bottom-up picture. In fact, by understanding the illness of the bulk theory, we can learn about the dysfunctional aspects of its boundary dual. This will be a main motivation for us, though we will see that many results we obtain can nevertheless be appreciated from a purely bulk perspective. 

The main character of this paper will be the shockwave in higher-spin theories. We will first present the shockwave solution in a general higher-spin theory and show that it solves all dynamical equations of motion exactly. This is a classical gravity result by itself, and we expect it to be useful even outside of the holographic context. The existence of the solution also does not rely on any specific sign of the cosmological constant, even though we assume it to be negative for the purpose of holographic interpretations. This is presented in detail in Section~\ref{sec:exact}. 

We then study the OTOC computable from the shockwave solution and derive the Lyapunov exponent $\li_L$ for a general theory containing fields with spins up to $\ell$. We find $\li_L=(\ell-1)2\pi T$, where $T$ is the temperature of the black hole. For $\ell\ge3$, the Lyapunov exponent exceeds the chaos bound \cite{Maldacena:2015waa}, suggesting that the dual CFT violates certain assumptions underlying the derivation of the bound, such as unitarity. We give the details in Section~\ref{sec:otoc}.

We also explore the relationship between the OTOC and a noteworthy feature of the retarded Green's function, known as pole skipping \cite{Grozdanov:2017ajz,Blake:2017ris,Blake:2018leo}. Intriguingly, the gravitational shockwave ($\ell=2$) can in fact be identified as a quasinormal mode at a special pole-skipping location \cite{Wang:2022mcq}. We generalize this to $\ell\ge2$ and use it to show that the leading pole-skipping point has frequency $\omega=\i \li_L$ and momentum $k=\i \li_L/v_B$ where $v_B$ is the butterfly velocity that appears in the OTOC computed from the shockwave. We present the technical details in Section~\ref{sec:ps}.

Next, we study whether the shockwave can manifest the causality issues of such higher-spin theories. We use a simple worldline approach to study the time delay when a probe particle goes across the shockwave. We discuss this in Section~\ref{sec:causal}.

We then end with a discussion of some open questions in Section~\ref{sec:disc}.

\section{Higher-spin shockwave as exact solution}\label{sec:exact}
Localized shockwaves are sourced along a one-dimensional worldline \cite{Roberts:2014isa}. Write the total sourced action as
\begin{equation}
    S_\text{total} = S + S_\text{source},
\end{equation}
where $S$ is a general diffeomorphism-invariant action in $d+2$ dimensions,
\begin{equation}\label{eq:action}
    S = \int \d^{d+2}x\sqrt{-g}\,\mathcal{L}\left(g,R,\nabla,\Phi\right),
\end{equation}
with $\mathcal{L}$ constructed out of the metric $g_{\mu\nu}$, the Riemann curvature tensor $R_{\mu\nu\rho\sigma}$, the covariant derivative operator $\nabla_\mu$, and a finite number of matter fields collectively denoted by $\Phi$, and $S_{\rm source}$ being a source term for the shockwave, whose specific form will be introduced later in a worldline formalism. 

For every field of spin $n$, define the equation of motion as
\begin{equation}\label{eq:eom_def}
    E^{(X)}_{\mu_1\dots \mu_n} \equiv \frac{1}{\sqrt{-g}} \frac{\delta S}{\delta X^{\mu_1 \dots \mu_n}}= -\frac{1}{\sqrt{-g}} \frac{\delta S_\text{source}}{\delta X^{\mu_1 \dots \mu_n}}\equiv T^{(X)}_{\mu_1\dots \mu_n}.
\end{equation}
We will refer to $T^{(X)}$ as the stress tensor even when $X$ is not the metric. As we will see, for the shockwave, $T^{(X)}$ is zero except for a specific component.

Suppose our highest-spin field has spin $\ell$ and for simplicity suppose there is only one such field which we will denote by $\phi_{\mu_1\dots\mu_\ell}$. The following formal derivation is insensitive to whether it is massive or massless, though we expect there may be qualitative differences when working with explicit examples. Suppose also that the theory admits a stationary planar black hole supported by stationary matter fields. The metric for such a solution can be written in Kruskal–Szekeres coordinates as
\bal \label{eq:metricKS}
\d s^2=2A(UV) \d U \d V+B(UV) \d y^i \d y^i,
\eal
where $U$ and $V$ both increase to the future and evaluate to zero on each of the two horizons respectively. The exact details of the functions $A(UV)$ and $B(UV)$ will depend on the theory and the matter profile, which we also assume to be stationary, isotropic, and homogeneous in $y^i$. When evaluating the functions on the horizon, we will denote $A_n\equiv \d^n A(UV)/\d (UV)^n|_{V=0}$ and similarly for $B_n$. By a rescaling of $y^i$, we fix $B_0=1$. Just like the metric, we also assume matter fields $\Phi$ are all smooth. In our coordinate system, this in particular means that $A(UV)$, $B(UV)$, and all components of matter fields are regular at the horizons.

To construct a shockwave solution, motivated by the form of the gravitational shockwave \cite{Roberts:2014isa}, which works for general higher-derivative gravity \cite{Dong:2022ucb}, we start with the following ansatz for a perturbation to the highest-spin field $\phi$:
\bal\label{eq:phiansatz}
\di \phi_{V(\ell)}\equiv \di \phi\underbrace{{}_{V\dots V}}_{\ell}=-{\ell}A_0^{\ell-1} h(y) \di (V),
\eal
where we used the notation $V(\ell)$ as a shorthand for $\ell$ instances of $V$. This perturbation vanishes everywhere except at $V=0$. The constants could be absorbed into the definition of $h(y)$ if desired. We will be studying perturbations to the equations of motion with $\delta \phi_{V(\ell)}$ as the perturbation parameter.

Now consider equations of motion. At this point, we will not distinguish between metric equations of motion and matter equations of motion. All that matters technically is the index structure. For simplicity, let us lower all indices using the metric. The zeroth-order equations of motion are satisfied by the assumption that \eqref{eq:metricKS} is a solution, so we start at the linear order. Consider a linearized equation of motion with $p$ instances of $U$-indices, $q$ instances of $V$-indices and an arbitrary number of $i$-indices. Denoting such a component by $\di E_{U(p),V(q)}$, where the positions of the indices are not specified, it must take the form
\bal
\di E_{U(p),V(q)}=\sum_{k\ge0} F_{k}(g, \Phi,\pd_V,\pd_U) \pd_V^k \di \phi_{V(\ell)},
\eal
where $F_k$ is a function constructed from background fields $A(UV)$, $B(UV)$, $\Phi$, and derivatives of them ($i$-derivatives are implicit as they are less relevant at this stage). Only this particular component of the highest-spin field appears on the right-hand side because all other fields and components are not perturbed in our ansatz. The perturbation does not have any $U$-dependence, so $\partial_U$ acting on $\delta\phi_{V(\ell)}$ would vanish.

Under a boost transformation ($U\to a\,U$, $V\to V/a$), the left-hand side transforms by a factor of $a^{q-p}$, whereas $\pd_V^k \di \phi_{V(\ell)}$ transforms by a factor of $a^{\ell+k}$. This then requires $F_k$ to behave as
\bal
F_k =  \sum_{m\ge0} \tilde{F}_{k,m}U^{m} V^{m+p-q+k+\ell},
\eal
for $\tilde{F}_{k,m}$ that are functions of $y^i$ only, and $m\ge0$ by the smoothness condition. Consequently, with some rewriting,
\bal
\di E_{U(p),V(q)}= \sum_{k,m\ge0} \tilde{F}'_{k,m}U^m V^{m+p-q+\ell} \(V \pd_V\)^k \di (V),
\eal
where $\tilde{F}'_{k,m}$ are another set of functions of $y^i$. This vanishes unless $m+p-q+\ell\le0$, which is only possible for $p=0$, $q=\ell$, i.e., when all indices are taken to be $V$, and when $m=0$. In other words, all the linearized equations of motion are automatically satisfied by \eqref{eq:phiansatz} except one: $\delta E_{V(\ell)}$. We will refer to it as the \emph{leading} equation of motion. For a localized shockwave with a point source, we add the following stress tensor
\begin{equation}\label{eq:source_gen}
    \delta T_{V(\ell)} = 
    T_0\,
    \delta^d(y) \delta(V),
\end{equation}
where $T_0$ is a constant.

Because the background fields are isotropic and homogeneous in $y^i$, the leading equation of motion must take the following form on the horizon:
\begin{equation}\label{eq:h_ode_gen}
    \sum_{n=0}^{n_{\text{max}}} c_n \(\delta^{ij}\partial_{i}\partial_{j}\)^n h(y) = 
    T_0\,\delta^d(y),
\end{equation} 
for some constants $c_n$. This is a differential equation for $h(y)$. The details of the solution depend on the $c_n$'s, which depend on the theory. If the field $\phi$ does not have higher-derivative interactions in the Lagrangian, $n_\text{max}=1$ and the differential equation can be solved exactly. In the presence of higher derivatives, we can treat the higher-derivative coupling constants perturbatively as done in the spin-2 case \cite{Mezei:2016wfz,Dong:2022ucb}. Corrections to $h(y)$ can be solved by substituting the zeroth-order solution and solving the perturbative equations order by order. 

Denoting the $n$-th order perturbation of the equation of motion $E$ by $\delta^{(n)}E$, we can similarly deduce that
\bal
\di^{(n)} E_{U(p),V(q)}
&= \sum_{k_1,\dots,k_n,m\ge0} \tilde{F}'_{k_1,\dots k_n,m} U^m V^{m+p-q+\ell n} \(V \pd_V\)^{k_1} \di (V)...\(V \pd_V\)^{k_n} \di (V)
\eal
where $\tilde{F}'_{k_1,\dots k_n,m}$ is yet another set of functions of $y^i$. It vanishes as a distribution when
\bal\label{eq:Idef}
I\equiv m+p-q+\ell n-n+1>0.
\eal
Recall that $\ell$ is the highest spin in the theory, so the following parameter, which counts indices, must be positive:
\bal
\Di\equiv \ell-(q-p)\geq 0.
\eal
We can then rewrite \eqref{eq:Idef} as
\bal
I=m+(\ell-1)(n-1)+\Di>0.
\eal
For $n=1$, i.e., at the linear order, we recover the fact that only (the $m=0$ part of) the leading equation of motion (the one with $\Di=0$) needs to be solved.

At higher orders ($n\ge2$), we see that $\ell\ge2$ would ensure that $I>0$ for all $\Di$ (regardless of $m$), i.e., all equations of motion are automatically satisfied by the ansatz \eqref{eq:phiansatz}. The shockwaves are therefore exact solutions. This equation also explains why shockwave solutions have not been found for scalar fields or vector fields: When $\ell=0$, higher-order equations of motion diverge; when $\ell=1$, the $\Delta=0$ equations of motion are non-vanishing at all orders, and there is no obvious solution. 

The standard gravitational shockwave has $\ell=2$ because the graviton has spin two. In this case, the same argument has been used to show that the gravitational shockwave solution exists as exact solutions in general higher-derivative gravity (without higher-spin fields) \cite{Dong:2022ucb}. 

For $\ell\ge3$, we call them higher-spin localized shockwaves. In this case, the metric remains smooth, while the highest-spin field $\phi$ has a distributional configuration. In this sense, one could say that these shockwaves are non-geometric.

As an example, consider the following higher-spin theory:
\bal
\(-\nabla^\nu \nabla_\nu+M^2\)\phi_{\mu_1...\mu_\ell}=0,~~~\nabla^\nu \phi_{\nu \mu_1...\mu_{\ell-1}}=0,~~~\phi^\nu_{~\nu\mu_1...\mu_{\ell-2}}=0.
\eal
On the background \eqref{eq:metricKS} and supposing $\phi$ vanishes on the background ($\phi=0+\delta \phi=\delta \phi$), the leading equation of motion is given by
\begin{align}
    0&=\(-A^{-1}\nabla_V \nabla_U-A^{-1}\nabla_U \nabla_V-B^{-1} \nabla_i\nabla_i + M^2\)\phi_{V(\ell)}\\
    &=\ell A^{-1}\pd_U \(\frac{UA^\pr}{A}\) \phi_{V(\ell)}-B^{-1}\pd_i\pd_i\phi_{V(\ell)}-\frac{d}{2}\frac{V B^\pr}{AB}\(\pd_V-\ell\frac{UA^\pr}{A}\)\phi_{V(\ell)}+M^2\phi_{V(\ell)}.\nonumber
\end{align}
Once we plug in \eqref{eq:phiansatz}, it simplifies to
\bal
\[-\frac{1}{B}\di^{ij}\frac{\pd}{\pd y^i}\frac{\pd}{\pd y^j}+\frac{d}{2}\frac{ B^\pr}{AB}+\ell \frac{A^\pr}{A^2} +M^2\]\phi_{V(\ell)}=0.
\eal
Adding a point-like source $T_{V(\ell)} = T_0\, \di(y)\di(V)$, we obtain
\bal
(\pd_i\pd_i -\mu^2) h(y) = \frac{T_0}{\ell A_0^{\ell-1}}\, \delta(y),
\eal
where
\bal
\mu^2=\frac{d}{2}\frac{ B_1}{A_0}+\ell \frac{A_1}{A_0^2} +M^2.
\eal
This can be solved exactly:
\bal
h(y) = -\frac{T_0}{\ell A_0^{\ell-1}} (2 \pi)^{-\frac{d}{2}}\(\frac{\mu}{|y|}\)^{\frac{d-2}{2}}K_{\frac{d-2}{2}}(\mu |y|),
\eal
where $K_\ai(z)$ is the modified Bessel function of the second kind. One can explicitly check that all other equations of motion vanish identically.

\section{Higher-spin shockwave as OTOC}
\label{sec:otoc}

In Einstein gravity, a classical derivation using the gravitational shockwave geometry gives the OTOC \cite{Shenker:2014cwa}. In this section, we show that the same derivation can be performed with higher spins, where the calculation reduces to the evaluation of the classical action of the higher-spin shockwave. 

Like in the case of (spin-2) gravity, the higher-spin shockwave is sourced by a localized stress tensor. Previously in \eqref{eq:source_gen}, the general form for the source was given as an assumption; we now derive its form using a worldline formalism.

Consider the following worldline action of a particle traveling along $X(\li)$ coupled to the higher-spin field $\phi$:
\bal\label{eq:source_action_hs}
S_\text{source}=\frac{c}{\ell}\int_{X(\li)} 
\frac{\d \li}{e^{\ell-1}}\,
\phi_{\mu_1...\mu_\ell}(X(\li)) \frac{\d X^{\mu_1}(\li)}{\d \li} \cdots\frac{\d X^{\mu_\ell}(\li)}{\d \li},
\eal
where $c$ indicates the strength of the coupling, whose sign is not fixed at this point. This can also be written as a spacetime integral by inserting appropriate delta functions. In particular, if the particle is localized at $y=0$ and travels along the horizon $V=0$, we can write it as
\begin{align}
    S_\text{source}
&=\frac{c}{\ell}\int  \d V \d^{d} y \,\delta(V)\delta^d(y)\int_{X(\li)} 
\frac{\d \li}{e^{\ell-1}}\,
\phi_{\mu_1...\mu_\ell} \frac{\d X^{\mu_1}}{\d \li} \cdots\frac{\d X^{\mu_\ell}}{\d \li}\nn
&=\frac{c}{\ell}\int \d U \d V \d^{d} y\frac{1}{e^{\ell-1}}  \frac{\d \li}{\d U} \, \di(V)\delta^d(y) \,\phi_{U...U} \(\frac{\d U}{\d \li}\)^\ell
\nn
&=\frac{c}{\ell}\int d^{d+2}x \frac{\sqrt{-g}}{A_0}\phi_{U...U}\,\di(V)\delta^d(y) \(\frac{1}{e}\frac{\d U}{\d \li}\)^{\ell-1}.
\end{align}
The non-zero component of the higher-spin stress tensor follows from this expression and scales as
\bal\label{eq:TVVV}
T_{VV\dots V}
&=-\frac{1}{\sqrt{-g}}\frac{\delta S_\text{source}}{\delta \phi^{VV\dots V}}
=-g_{UV}^{\ell}\frac{1}{\sqrt{-g}}\frac{\delta S_\text{source}}{\delta \phi_{UU\dots U}}
\nn
&=
-\frac{c}{\ell}A_0^{\ell-1}\di(V)\delta^d(y) \(\frac{1}{e}\frac{\d U}{\d \li}\)^{\ell-1}
=-\frac{c}{\ell}{A_0^{\ell-1}}\di(V)\delta^d(y) \(p_1^U\)^{\ell-1},
\eal
where $p_1$ is the momentum of the particle that generates the shockwave under consideration. There is another similar shockwave solution generated by a particle with momentum $p_2$ along the other horizon. We now follow \cite{Shenker:2014cwa} to find the Lyapunov exponent of the theory. 

With this source, the leading equation of motion will lead to the component $\phi_{V(\ell)}$ taking the form
\begin{equation}\label{eq:phiVVV}
    \phi_{V...V}\propto A_0^{\ell-1}\(p_1^U\)^{\ell-1} \di (V).
\end{equation}
To find the Lyapunov exponent, use the fact that the scattering between two particles sourcing the shockwave is given by
\begin{equation}
    e^{\i \di}=e^{\i S_\text{total}},
\end{equation}
where $S_\text{total}$ is the action of the backreacted classical solution which, upon substitution of \eqref{eq:TVVV} and \eqref{eq:phiVVV}, scales as
\bal
S_\text{total}
\sim  \int \sqrt{-g}\,\phi_{\mu_1\dots \mu_\ell}T^{\mu_1\dots \mu_\ell}
\propto (A_0 p_1^Up_2^V)^{\ell-1}.
\eal
The center-of-mass energy, $s$, is related to these quantities by $s=-2A_0p_1^{U}p_2^V$. Recalling that the center-of-mass energy scales as the relative boost $s\sim e^{2\pi T t}$, the OTOC is order unity when
\begin{equation}
    \di\sim1 \implies e^{(\ell-1)2\pi T t} \sim 1.
\end{equation}
The Lyapunov exponent is therefore given by
\bal\label{eq:h_largey}
\li_L= (\ell-1)2\pi T.
\eal
This agrees with the prediction from a pole-skipping analysis \cite{Wang:2022mcq}. In the massless case, this implies that holographic CFTs with finitely many conserved currents violate the chaos bound of \cite{Maldacena:2015waa}, a result that was already derived in AdS${}_3$/CFT${}_2$ in \cite{Perlmutter:2016pkf}. In the massive case, this result suggests that unitary CFTs with a finite tower of single-trace higher-spin local operators of finite conformal dimensions at large $N$ cannot have a weakly coupled local bulk dual, consistent with \cite{Heemskerk:2009pn}.

So far, the $y$ dependence has been suppressed in the above formulas. Restoring it leads to a factor of $h(y)$ in $S_\text{total}$. For large $|y|$, 
\bal
    h(y) \sim e^{-\mu |y|},
\eal
so the OTOC is of order unity when
\begin{equation}
    e^{(\ell-1)2\pi T t-\mu |y|} \sim 1.
\end{equation}
From this, it is clear that the butterfly velocity is given by
\begin{equation}
    v_B^{(\ell)} = \frac{(\ell-1)2\pi T}{\mu}.
\end{equation}

\section{Higher-spin shockwave as skipped pole}
\label{sec:ps}

In thermal retarded Green's functions, it can happen that the residue of a pole is zero. Such a pole is referred to as \emph{skipped}. The division of zero by zero at such a point makes the Green's function ill-defined: The value of the Green's function at such a pole-skipping point depends on how the limit is taken in the space of (complex) frequency and momenta. For holographic systems, this multi-valuedness or ambiguity turns out to have an interesting geometric realization. 

Pole skipping was first discovered in Einstein gravity, in the Green's function of a certain channel of the stress tensor, at frequency $\oi$ and momentum $k$ given by
\begin{equation}
    \omega=\i 2\pi T, \quad k=\i 2\pi T/v_B,
\end{equation}
where $v_B$ is the butterfly velocity of the boundary CFT theory dual to Einstein gravity. This was a particularly important one as it is related to quantities that characterize and quantify the chaotic properties of the corresponding boundary theory.

The underlying technical reason why the holographic Green's function and OTOC both know about the butterfly velocity boils down to the fact that the quasinormal mode at the special pole-skipping point \emph{is} the shockwave. For gravitational theories without higher spins, this was argued in \cite{Wang:2022mcq} where it was shown that a localized gravitational shockwave along the black hole horizon can be obtained from the special quasinormal mode after some appropriate regularization and limiting procedure. We now show that the same argument can be used in the presence of higher spins.

To begin with, recall that a theory with spins up to $\ell$ has its first pole-skipping point located at $\omega=\i (\ell-1)2\pi T$, derived for general integer $\ell$ in \cite{Wang:2022mcq} whose notations and terminologies we will follow in this section. 

It is useful to switch to Eddington-Finkelstein coordinates, where the metric takes the form
\begin{align}
\d s^2=-f(r) \d v^2+2 \d v \d r+h(r) \d x^i \d x^i.
\end{align}
The horizon is at $r=r_0$, and the temperature is given by $2\pi T=f'(r_0)/2$. It is related to \eqref{eq:metricKS} via
\begin{align}
U=-e^{-f^{\prime}\left(r_0\right)\left(v-2 r_*\right) / 2}, \quad V=e^{f^{\prime}\left(r_0\right) v / 2}, \quad \d r_* / \d r=1 / f(r).
\end{align}

Substituting the frequency $\omega=\mathrm{i}(\ell-1) 2 \pi T$ into the expression for a Fourier mode, $e^{-\i \omega v}$, we have
\begin{equation}
    \di \phi_{v(\ell)}\sim e^{-\i \omega v}= e^{(\ell-1)2\pi T\frac{2}{f^{\prime}\left(r_0\right)} \log V}=V^{\ell-1}.
\end{equation}
In Kruskal–Szekeres coordinates, the relevant component is then given by
\begin{equation}
    \di \phi_{V(\ell)} = \left(\frac{\pd v}{\pd V}\right)^\ell \di \phi_{v(\ell)}=\left(\frac{2}{f'(r_0)V}\right)^\ell \di \phi_{v(\ell)} =
    \left(\frac{1}{2\pi TV}\right)^\ell \di \phi_{v(\ell)} 
    \sim \frac{1}{V}.
\end{equation}
The highest-weight equation of motion
\beq
\di E_{v(\ell)}=\sum_{k,l} H_{k,l}(f,h,\pd_r,\Phi)(\pd_v)^{k}(\pd_i)^{l} \di \phi_{v(\ell)},
\label{eq:Erv}
\eeq
after using
\beq
\di E_{V(\ell)}=\left(\frac{1}{2\pi TV}\right)^\ell \di E_{v(\ell)}+\cdots,
\label{eq:map}
\eeq
where the dots represent subleading corrections that vanish on the horizon, transforms to
\bal
\label{eq:big_EUV}
\di E_{V(\ell)}&=\left(\frac{1}{2\pi T}\right)^\ell \frac{1}{V^\ell}\sum_{k,m} H_{k,m}(\pd_v)^{k}(\pd_i)^{m} \di \phi_{v(\ell)}\nn
&=\left(\frac{1}{2\pi T}\right)^\ell \frac{1}{V^\ell} \sum_{k,m} H_{k,m}(\pd_i)^{m}\(\frac{1}{2\pi T}V\pd_V\)^{k} \di \phi_{v(\ell)}\nn
&=\left(\frac{1}{2\pi T}\right)^\ell \sum_{k,m} H_{k,m}(\pd_i)^{m}\(\frac{1}{2\pi T}(V\pd_V+\ell)\)^{k}\frac{ \di \phi_{v(\ell)} }{V^\ell}\nn
&=\sum_{k,m} \tilde{H}_{k,m}(\pd_i)^{m}\(V\pd_V\)^{k}\left[\left(\frac{1}{2\pi T}\right)^\ell\di \phi_{v(\ell)}\right]\nn
&=\sum_{k,m} \tilde{H}_{k,m}(\pd_i)^{m}\(V\pd_V\)^{k} \di \phi_{V(\ell)}.
\eal
Now compare
\begin{equation}
    \delta \phi_{V(\ell)} \sim \frac{1}{V}\, e^{\i k y}
\end{equation}
with the shockwave perturbation which for large $y$ looks like
\begin{equation}
    \delta \phi_{V(\ell)} \sim \di(V)\, e^{-\mu |y|}.
\end{equation}
As pointed out in \cite{Grozdanov:2017ajz}, $1/V$ has the same distributional properties as $\delta V$ under the operator $V\partial_V$, so replacing $1/V$ with $\di V$ in the special quasinormal mode solution leads to a new solution of the equations of motion, at least at the linearized level. We identify this with the shockwave solution presented in Sec.~\ref{sec:exact}. The fact that it turns out to be an exact solution follows from the rest of that section. For very large $|y|$, the two modes locally look the same, so we can identify $k=\i \mu$.\footnote{One can also study pole skipping in a spherical basis rather than the plane wave basis, which makes the connection to the shockwave more precise \cite{Chua:2025vig}.} It then follows that the butterfly velocity defined from pole skipping is the same as that defined using the shockwave.

\section{Higher-spin shockwave as time machine}
\label{sec:causal}

Let us now discuss issues related to causality. It is well-known that isolated higher-spin fields are causally pathological \cite{Camanho:2014apa}. We will see in this section how this might be manifested in the shockwave. 

Before discussing higher spins, let us review what happens when a probe particle crosses the (spin-2) gravitational shockwave. Following \eqref{eq:phiansatz}, pick the metric perturbation $\di g_{VV}= -2 A_0\,h(y)\di(V)$. Take a probe particle $X^\mu(\li)$ with worldline action
\be
S_0[X(\li)]=\int \frac{\d \li}{e} \, \frac{1}{2} g_{\mu\nu}(X(\li)) \frac{\d X^\mu}{\d \li}\frac{\d X^\nu}{\d \li}.
\ee
Its equation of motion is then given by the geodesic equation
\begin{equation}
\frac{\d^2 X^\mu}{\d \li^2} + \(g^{\mu\nu}\pd_\ri g_{\nu\si}-\frac{1}{2}\pd^\mu g_{\ri\si}\) \frac{\d X^\ri}{\d \li} \frac{\d X^\si}{\d \li}=0.
\end{equation}
From this, one can check that $X^V$ is an affine parameter. The $U$ component gives
\bal
\frac{\d^2 X^U}{\d V^2} -h(y)\pd_V \di(V)=0,
\eal
which shows that $X^U$ is shifted by
\bal
\int_{-\epi}^{\epi} \d V\frac{\d X^U}{\d V}=h(y),
\eal
i.e., the probe particle jumps along the $U$ direction by an amount of $h(y)$ upon crossing the shockwave. In Einstein gravity, $h(y)$ is positive \cite{Roberts:2014isa}; higher-derivative corrections are treated perturbatively, so $h(y)$ remains positive. This means that the probe particle experiences a time delay rather than an advance. Causality is respected. 

Now consider a simple spin-3 example $\phi_{\mu\nu\rho}$ that is totally symmetric in the indices. Take the shockwave solution, i.e., \eqref{eq:phiansatz} with $\ell=3$ and consider a simple effective worldline action that couples the probe particle to both gravity and the spin-3 field:
\bal
S_\text{probe}[X(\li)]=S_0+\frac{c}{3} \int \frac{\d \li}{e^2}\,\phi_{\mu\nu\ri}(X(\li)) \frac{\d X^\mu}{\d \li} \frac{\d X^\nu}{\d \li} \frac{\d X^\ri}{\d \li}.
\eal
The equation of motion for the probe particle is given by
\bal
&\frac{\d^2 X^\mu}{\d \li^2} + \(g^{\mu\nu}\pd_\ri g_{\nu\si}-\frac{1}{2}\pd^\mu g_{\ri\si}\) \frac{\d X^\ri}{\d \li} \frac{\d X^\si}{\d \li}\nn
+\,&c\, g^{\mu\ti}\(2\phi_{\ti\nu\ri}\frac{\d^2 X^\nu}{\d \li^2}\frac{\d X^\ri}{\d \li} + \(\pd_\si \phi_{\ti\nu\ri}-\frac{1}{3}\pd_\ti\phi_{\nu\ri\si}\) \frac{\d X^\nu}{\d \li}\frac{\d X^\ri}{\d \li}\frac{\d X^\si}{\d \li}\)=0.
\eal
Again, $X^V$ is an affine parameter. Using \eqref{eq:phiansatz}, the $U$ component gives
\bal
\frac{\d^2 X^U}{\d V^2}- 2c A_0h(y) \pd_V\di(V)=0,
\eal
which gives us a shift of $2cA_0h(y)$. 

More generally, in the (totally symmetric) spin-$\ell$ case,
\bal
S_\text{probe}[X(\li)]=S_0+\frac{c}{\ell} \int \frac{\d \li}{e^{\ell-1}}\,\phi_{\mu_1\mu_2\dots\mu_\ell}(X(\li)) \frac{\d X^{\mu_1}}{\d \li} \frac{\d X^{\mu_2}}{\d \li}\cdots \frac{\d X^{\mu_\ell}}{\d \li},
\eal
and the worldline equation of motion leads to a shift of
\begin{equation}
    \text{shift} = (\ell-1)cA_0^{\ell-2}h(y)
\end{equation}
for the shockwave \eqref{eq:phiansatz}. Note that we have taken the second term in the probe particle worldline action to have the same form as the particle that sources the shockwave \eqref{eq:source_action_hs}.

To figure out whether the probe particle experiences a delay or an advance in this case, we need to know the sign of $h(y)$ relative to $c$. Solving the differential equation \eqref{eq:h_ode_gen} at leading order (i.e., without higher derivatives), one can explicitly show that $h(y)$ has the same sign (for all $y$) as $-T_0/c_2$. From \eqref{eq:TVVV}, $T_0$ in \eqref{eq:h_ode_gen} has the same sign as $-cA_0^{\ell-1}$. The shift therefore has the same sign as $-c_2$. If the shift is positive, the shockwave acts like a time machine that only sends the particle to the future, which is allowed by causality; if the shift is negative, we can take two such shockwaves and form a closed timeline curve, thus making a time machine that violates causality. This serves as a simple test of causality. In Einstein gravity, $c_2<0$, so causality is obeyed. 

Moreover, there is also a purely boundary notion of causality that we can test using the shockwave. We saw previously how the shockwave computes the butterfly velocity of the boundary theory. Now, if the butterfly velocity is larger than the boundary speed of light, then the boundary CFT is acausal. For the example in Sec.~\ref{sec:exact},
\begin{equation}
    v_B^{(\ell)}=\frac{(\ell-1)2\pi T}{\sqrt{\frac{d}{2}\frac{ B_1}{A_0}+\ell \frac{A_1}{A_0^2} +M^2}}.
\end{equation}
For large $\ell$, this scales as $\sqrt{\ell}$, so it will exceed the speed of light for large enough $\ell$. 

\section{Discussion}
\label{sec:disc}

By utilizing an argument based on the boost symmetry of the black hole, we have shown that shockwaves generally exist as classical solutions in gravitational theories with a local, diffeomorphism-invariant action, even when higher-spin fields are present. In this context, it is the highest-spin field, rather than the metric, that exhibits a Dirac delta function supported at the horizon. With this new tool, we see that various features of quantum chaos in AdS/CFT naturally generalize to higher spins.

In particular, we have used higher-spin shockwaves in AdS${}_{d+2}$ to argue that the dual CFT${}_{d+1}$ would violate the chaos bound, complementing a result in CFT${}_2$ \cite{Perlmutter:2016pkf}. This in turn implies that the CFT would violate causality \cite{Hartman:2015lfa}. In the massless case, this also provides an alternative understanding of why higher-spin conserved currents in holographic CFTs are disallowed \cite{Maldacena:2011jn,Boulanger:2013zza,Alba:2015upa,Hartman:2015lfa,Afkhami-Jeddi:2017idc}.

An interesting subtlety to highlight is that, although we have argued that the special quasinormal mode corresponding to the leading-order pole skipping is equivalent to the shockwave, the derivations of the Lyapunov exponent in the pole-skipping and shockwave calculations differ. In other words, the two calculations appear independent, making their agreement nontrivial. In contrast, the two butterfly velocities share the same origin, as we have demonstrated that one calculation reduces to the other.

One possible route to better understand the connection between the shockwave and the pole-skipping mode is through horizon symmetries. A useful first step would be to investigate if the connection of horizon symmetries to shockwaves and pole skipping studied in \cite{Knysh:2024asf} generalizes to the higher-spin case. This might shed light on why two seemingly independent calculations of the Lyapunov exponent agree.

It is also interesting to note that imposing higher-spin versions of the averaged null energy condition \cite{Hartman:2016lgu,Kravchuk:2018htv,Meltzer:2018tnm} do not seem to affect the causal property of the shockwave. This is because the sign of the shift is independent of the sign of the stress tensor; instead, we found it to depend solely on a specific sign of the kinetic term. However, our analysis is subject to several limitations. For example, we have treated higher-derivative terms in the action perturbatively in order to solve the differential equation for the function $h(y)$, but the Lagrangian for higher-spin fields is highly constrained and may not allow such coupling constants to be arbitrary. Moreover, we have chosen a specific form of the worldline action, which is not the most general. We anticipate that a detailed analysis of gauge symmetries and the index structure of kinetic terms for higher-spin fields could yield stronger conclusions on this causality issue. One could also try to systematically derive the conditions under which causality violation can be avoided, though it is likely that they would be unphysical.

One might question our assumption of the existence of a stationary black hole solution in a general theory. Should the bulk theory not admit such a solution, the boundary theory is expected to be integrable or at least sub-maximally chaotic (in the large-$N$ limit), given that black holes are the fastest scramblers \cite{Hayden:2007cs,Sekino:2008he}. In this case, our shockwave solutions also do not exist, so we cannot calculate the OTOC this way.

We have focused on isolated higher spins. One step that leverages this restriction is our ansatz, which presumes the existence of a finite highest spin. While this may seem like a mere technical convenience, the fact that there are higher-spin gravity theories with infinitely many spins dual to free boundary theories ($\lambda_L=0$) \cite{Klebanov:2002ja} suggests that this assumption is essential. More conjecturally, the existence of the localized shockwave solution on a black hole background in the bulk theory might be an essential ingredient for the boundary theory to be chaotic, a statement that would be interesting to refine.

On the other hand, we imposed a related restriction that there is only one field with the highest spin. This can most likely be lifted, though it would require a more careful analysis of the differential equations resulting from the highest-weight components of the equations of motion, now more than one. 

Our study is limited to bosonic fields. The pole-skipping analysis, on the other hand, has been performed with both bosonic and fermionic fields, with the leading pole-skipping frequency given by $\omega=\i(\ell-1)2\pi T$ for both integer and half-integer $\ell$ \cite{Ceplak:2019ymw,Ceplak:2021efc,Ning:2023ggs}. It would be interesting to explore whether an analogous fermionic shockwave exists that can be used to derive $\lambda_L=(\ell-1)2\pi T$ for half-integer values of $\ell$.

The higher-spin shockwaves we studied, though non-geometric, propagate on black hole backgrounds, which are geometric. One may then ask: are there shockwaves that propagate on higher-spin black hole backgrounds \cite{Gutperle:2011kf,Ammon:2011nk}, which are themselves non-geometric? It would be interesting to investigate the holographic consequences of such geometries, if they exist. Moreover, as the notion of a black hole can become gauge-dependent, one can ask how to reproduce the OTOC that is easily computed in the black hole gauge if a gauge is chosen such that the spacetime is e.g.~a traversable wormhole \cite{Ammon:2011nk}. Relatedly, one can ask what becomes of the shockwave under such gauge transformations.

As far as the shockwave solutions are concerned, we did not have to distinguish between massive and massless fields. However, since they are physically very different, it is likely that a careful study of the properties of the shockwaves would reveal a distinction. We also did not restrict the spacetime dimension to be higher than three. It is possible that pure higher-spin gravity in three bulk dimensions would require special treatments which may render our argument invalid, given that such theories have no local degrees of freedom. For three-dimensional theories that do possess local degrees of freedom, we expect them to closely resemble the higher-dimensional case.

The agreement between the two butterfly velocities goes beyond geometry because, as we have seen, the key component in the calculation is a component of the highest-spin field, which is not always the metric. The same cannot be said about another definition of butterfly velocity, one that makes use of the RT surface \cite{Mezei:2016wfz}. With strong evidence, this butterfly velocity is believed to agree with the other two in general higher-derivative gravity without matter (or with low-spin matter only) \cite{Dong:2022ucb}. We can now ask whether this definition can be generalized to allow higher-spin fields and, if so, whether the butterfly velocity would still agree. An obstacle to this generalization is that there are difficulties in generalizing the derivation of the RT formula \`a la \cite{Lewkowycz:2013nqa,Dong:2013qoa} in the presence of higher-spin fields.\footnote{In \cite{Dong:2023bax}, the problem is avoided through a large-mass expansion, but a general solution remains unknown. Issues are also encountered in generalizing the Wall entropy \cite{Wall:2015raa} to higher spins \cite{Yan:2024gbz}.} It is likely that the ``dominance'' of the graviton over other fields is a fundamental requirement for the boundary entropy to have a classical bulk manifestation.  It would be interesting to either prove or disprove this assertion.

In AdS/CFT, bulk renormalization group flow is immaterial from the boundary perspective, so it does not matter whether a certain quantity is computed in the UV or the IR. This serves as a simple tool for verifying the consistency of proposed bulk duals for certain CFT quantities. For instance, the RT formula passes this test \cite{Dong:2023bax}. The same reasoning applies to quantities such as the Lyapunov exponent and the butterfly velocity. Contrary to this expectation, if we take in the UV the spin-4 example in \cite{Dong:2023bax} which flows to a purely metric theory (spin-2), the shockwave analysis appears to predict a Lyapunov exponent of $6\pi T$ in the UV but $2\pi T$ in the IR. In this particular example, the resolution lies in the fact that this spin-4 field can have at most two (lower) $V$-indices due to symmetry constraints, rendering the shockwave ansatz inapplicable. In this sense, it is a fake spin-4 field. In situations where the shockwave analysis is valid in the UV, it remains unclear how to reconcile this discrepancy, and we leave its resolution to future work.

Finally, it would be interesting to investigate higher-spin shockwaves on extremal and cosmological horizons. 

\acknowledgments
It is a pleasure to thank Mike Blake, Tom Hartman, Matt Heydeman, Jingping Li, Julio Parra-Martinez, Eric Perlmutter, Amir Tajdini, Zixia Wei, Wayne Weng, Zihan Yan, and Xi Yin for helpful discussions. DW is supported by NSF grant PHY-2207659. ZYW is supported by funds from the University of California.



\bibliographystyle{JHEP}
\bibliography{bibliography}

\providecommand{\href}[2]{#2}\begingroup\raggedright\begin{thebibliography}{10}

\bibitem{Maldacena:1997re}
J.M.~Maldacena, \emph{{The Large N limit of superconformal field theories and supergravity}}, \href{https://doi.org/10.4310/ATMP.1998.v2.n2.a1}{\emph{Adv. Theor. Math. Phys.} {\bfseries 2} (1998) 231} [\href{https://arxiv.org/abs/hep-th/9711200}{{\ttfamily hep-th/9711200}}].

\bibitem{Witten:1998qj}
E.~Witten, \emph{{Anti-de Sitter space and holography}}, \href{https://doi.org/10.4310/ATMP.1998.v2.n2.a2}{\emph{Adv. Theor. Math. Phys.} {\bfseries 2} (1998) 253} [\href{https://arxiv.org/abs/hep-th/9802150}{{\ttfamily hep-th/9802150}}].

\bibitem{Ryu:2006bv}
S.~Ryu and T.~Takayanagi, \emph{{Holographic derivation of entanglement entropy from AdS/CFT}}, \href{https://doi.org/10.1103/PhysRevLett.96.181602}{\emph{Phys. Rev. Lett.} {\bfseries 96} (2006) 181602} [\href{https://arxiv.org/abs/hep-th/0603001}{{\ttfamily hep-th/0603001}}].

\bibitem{Ryu:2006ef}
S.~Ryu and T.~Takayanagi, \emph{{Aspects of Holographic Entanglement Entropy}}, \href{https://doi.org/10.1088/1126-6708/2006/08/045}{\emph{JHEP} {\bfseries 08} (2006) 045} [\href{https://arxiv.org/abs/hep-th/0605073}{{\ttfamily hep-th/0605073}}].

\bibitem{Shenker:2013pqa}
S.H.~Shenker and D.~Stanford, \emph{{Black holes and the butterfly effect}}, \href{https://doi.org/10.1007/JHEP03(2014)067}{\emph{JHEP} {\bfseries 03} (2014) 067} [\href{https://arxiv.org/abs/1306.0622}{{\ttfamily 1306.0622}}].

\bibitem{Roberts:2014isa}
D.A.~Roberts, D.~Stanford and L.~Susskind, \emph{{Localized shocks}}, \href{https://doi.org/10.1007/JHEP03(2015)051}{\emph{JHEP} {\bfseries 03} (2015) 051} [\href{https://arxiv.org/abs/1409.8180}{{\ttfamily 1409.8180}}].

\bibitem{Roberts:2016wdl}
D.A.~Roberts and B.~Swingle, \emph{{Lieb-Robinson Bound and the Butterfly Effect in Quantum Field Theories}}, \href{https://doi.org/10.1103/PhysRevLett.117.091602}{\emph{Phys. Rev. Lett.} {\bfseries 117} (2016) 091602} [\href{https://arxiv.org/abs/1603.09298}{{\ttfamily 1603.09298}}].

\bibitem{Shenker:2013yza}
S.H.~Shenker and D.~Stanford, \emph{{Multiple Shocks}}, \href{https://doi.org/10.1007/JHEP12(2014)046}{\emph{JHEP} {\bfseries 12} (2014) 046} [\href{https://arxiv.org/abs/1312.3296}{{\ttfamily 1312.3296}}].

\bibitem{Shenker:2014cwa}
S.H.~Shenker and D.~Stanford, \emph{{Stringy effects in scrambling}}, \href{https://doi.org/10.1007/JHEP05(2015)132}{\emph{JHEP} {\bfseries 05} (2015) 132} [\href{https://arxiv.org/abs/1412.6087}{{\ttfamily 1412.6087}}].

\bibitem{Aichelburg:1970dh}
P.C.~Aichelburg and R.U.~Sexl, \emph{{On the Gravitational field of a massless particle}}, \href{https://doi.org/10.1007/BF00758149}{\emph{Gen. Rel. Grav.} {\bfseries 2} (1971) 303}.

\bibitem{Dray:1984ha}
T.~Dray and G.~'t~Hooft, \emph{{The Gravitational Shock Wave of a Massless Particle}}, \href{https://doi.org/10.1016/0550-3213(85)90525-5}{\emph{Nucl. Phys.} {\bfseries B253} (1985) 173}.

\bibitem{Sfetsos:1994xa}
K.~Sfetsos, \emph{{On gravitational shock waves in curved space-times}}, \href{https://doi.org/10.1016/0550-3213(94)00573-W}{\emph{Nucl. Phys.} {\bfseries B436} (1995) 721} [\href{https://arxiv.org/abs/hep-th/9408169}{{\ttfamily hep-th/9408169}}].

\bibitem{Horowitz:1999gf}
G.T.~Horowitz and N.~Itzhaki, \emph{{Black holes, shock waves, and causality in the AdS / CFT correspondence}}, \href{https://doi.org/10.1088/1126-6708/1999/02/010}{\emph{JHEP} {\bfseries 02} (1999) 010} [\href{https://arxiv.org/abs/hep-th/9901012}{{\ttfamily hep-th/9901012}}].

\bibitem{Cornalba:2006xk}
L.~Cornalba, M.S.~Costa, J.~Penedones and R.~Schiappa, \emph{{Eikonal Approximation in AdS/CFT: From Shock Waves to Four-Point Functions}}, \href{https://doi.org/10.1088/1126-6708/2007/08/019}{\emph{JHEP} {\bfseries 08} (2007) 019} [\href{https://arxiv.org/abs/hep-th/0611122}{{\ttfamily hep-th/0611122}}].

\bibitem{Cornalba:2006xm}
L.~Cornalba, M.S.~Costa, J.~Penedones and R.~Schiappa, \emph{{Eikonal Approximation in AdS/CFT: Conformal Partial Waves and Finite N Four-Point Functions}}, \href{https://doi.org/10.1016/j.nuclphysb.2007.01.007}{\emph{Nucl. Phys. B} {\bfseries 767} (2007) 327} [\href{https://arxiv.org/abs/hep-th/0611123}{{\ttfamily hep-th/0611123}}].

\bibitem{Cornalba:2007zb}
L.~Cornalba, M.S.~Costa and J.~Penedones, \emph{{Eikonal approximation in AdS/CFT: Resumming the gravitational loop expansion}}, \href{https://doi.org/10.1088/1126-6708/2007/09/037}{\emph{JHEP} {\bfseries 09} (2007) 037} [\href{https://arxiv.org/abs/0707.0120}{{\ttfamily 0707.0120}}].

\bibitem{Nozaki:2013wia}
M.~Nozaki, T.~Numasawa and T.~Takayanagi, \emph{{Holographic Local Quenches and Entanglement Density}}, \href{https://doi.org/10.1007/JHEP05(2013)080}{\emph{JHEP} {\bfseries 05} (2013) 080} [\href{https://arxiv.org/abs/1302.5703}{{\ttfamily 1302.5703}}].

\bibitem{Camanho:2014apa}
X.O.~Camanho, J.D.~Edelstein, J.~Maldacena and A.~Zhiboedov, \emph{{Causality Constraints on Corrections to the Graviton Three-Point Coupling}}, \href{https://doi.org/10.1007/JHEP02(2016)020}{\emph{JHEP} {\bfseries 02} (2016) 020} [\href{https://arxiv.org/abs/1407.5597}{{\ttfamily 1407.5597}}].

\bibitem{Asplund:2014coa}
C.T.~Asplund, A.~Bernamonti, F.~Galli and T.~Hartman, \emph{{Holographic Entanglement Entropy from 2d CFT: Heavy States and Local Quenches}}, \href{https://doi.org/10.1007/JHEP02(2015)171}{\emph{JHEP} {\bfseries 02} (2015) 171} [\href{https://arxiv.org/abs/1410.1392}{{\ttfamily 1410.1392}}].

\bibitem{Hartman:2015lfa}
T.~Hartman, S.~Jain and S.~Kundu, \emph{{Causality Constraints in Conformal Field Theory}}, \href{https://doi.org/10.1007/JHEP05(2016)099}{\emph{JHEP} {\bfseries 05} (2016) 099} [\href{https://arxiv.org/abs/1509.00014}{{\ttfamily 1509.00014}}].

\bibitem{Afkhami-Jeddi:2016ntf}
N.~Afkhami-Jeddi, T.~Hartman, S.~Kundu and A.~Tajdini, \emph{{Einstein gravity 3-point functions from conformal field theory}}, \href{https://doi.org/10.1007/JHEP12(2017)049}{\emph{JHEP} {\bfseries 12} (2017) 049} [\href{https://arxiv.org/abs/1610.09378}{{\ttfamily 1610.09378}}].

\bibitem{Afkhami-Jeddi:2017rmx}
N.~Afkhami-Jeddi, T.~Hartman, S.~Kundu and A.~Tajdini, \emph{{Shockwaves from the Operator Product Expansion}}, \href{https://doi.org/10.1007/JHEP03(2019)201}{\emph{JHEP} {\bfseries 03} (2019) 201} [\href{https://arxiv.org/abs/1709.03597}{{\ttfamily 1709.03597}}].

\bibitem{Fitzpatrick:2019efk}
A.L.~Fitzpatrick, K.-W.~Huang and D.~Li, \emph{{Probing universalities in d \ensuremath{>} 2 CFTs: from black holes to shockwaves}}, \href{https://doi.org/10.1007/JHEP11(2019)139}{\emph{JHEP} {\bfseries 11} (2019) 139} [\href{https://arxiv.org/abs/1907.10810}{{\ttfamily 1907.10810}}].

\bibitem{Maldacena:2001kr}
J.M.~Maldacena, \emph{{Eternal black holes in anti-de Sitter}}, \href{https://doi.org/10.1088/1126-6708/2003/04/021}{\emph{JHEP} {\bfseries 04} (2003) 021} [\href{https://arxiv.org/abs/hep-th/0106112}{{\ttfamily hep-th/0106112}}].

\bibitem{Dong:2013qoa}
X.~Dong, \emph{{Holographic Entanglement Entropy for General Higher Derivative Gravity}}, \href{https://doi.org/10.1007/JHEP01(2014)044}{\emph{JHEP} {\bfseries 01} (2014) 044} [\href{https://arxiv.org/abs/1310.5713}{{\ttfamily 1310.5713}}].

\bibitem{upcomingRT}
X.~Dong, D.~Wang, W.W.~Weng and J.~Xu, \emph{Revisiting holographic entanglement entropy for general higher-derivative gravity}, {\emph{to appear} }.

\bibitem{Mezei:2016wfz}
M.~Mezei and D.~Stanford, \emph{{On entanglement spreading in chaotic systems}}, \href{https://doi.org/10.1007/JHEP05(2017)065}{\emph{JHEP} {\bfseries 05} (2017) 065} [\href{https://arxiv.org/abs/1608.05101}{{\ttfamily 1608.05101}}].

\bibitem{Alishahiha:2016cjk}
M.~Alishahiha, A.~Davody, A.~Naseh and S.F.~Taghavi, \emph{{On Butterfly effect in Higher Derivative Gravities}}, \href{https://doi.org/10.1007/JHEP11(2016)032}{\emph{JHEP} {\bfseries 11} (2016) 032} [\href{https://arxiv.org/abs/1610.02890}{{\ttfamily 1610.02890}}].

\bibitem{Qaemmaqami:2017bdn}
M.M.~Qaemmaqami, \emph{{Criticality in third order lovelock gravity and butterfly effect}}, \href{https://doi.org/10.1140/epjc/s10052-018-5541-6}{\emph{Eur. Phys. J. C} {\bfseries 78} (2018) 47} [\href{https://arxiv.org/abs/1705.05235}{{\ttfamily 1705.05235}}].

\bibitem{Li:2017nxh}
W.-J.~Li, P.~Liu and J.-P.~Wu, \emph{{Weyl corrections to diffusion and chaos in holography}}, \href{https://doi.org/10.1007/JHEP04(2018)115}{\emph{JHEP} {\bfseries 04} (2018) 115} [\href{https://arxiv.org/abs/1710.07896}{{\ttfamily 1710.07896}}].

\bibitem{Dong:2022ucb}
X.~Dong, D.~Wang, W.W.~Weng and C.-H.~Wu, \emph{{A tale of two butterflies: an exact equivalence in higher-derivative gravity}}, \href{https://doi.org/10.1007/JHEP10(2022)009}{\emph{JHEP} {\bfseries 10} (2022) 009} [\href{https://arxiv.org/abs/2203.06189}{{\ttfamily 2203.06189}}].

\bibitem{Bekaert:2010hw}
X.~Bekaert, N.~Boulanger and P.~Sundell, \emph{{How higher-spin gravity surpasses the spin two barrier: no-go theorems versus yes-go examples}}, \href{https://doi.org/10.1103/RevModPhys.84.987}{\emph{Rev. Mod. Phys.} {\bfseries 84} (2012) 987} [\href{https://arxiv.org/abs/1007.0435}{{\ttfamily 1007.0435}}].

\bibitem{Maldacena:2011jn}
J.~Maldacena and A.~Zhiboedov, \emph{{Constraining Conformal Field Theories with A Higher Spin Symmetry}}, \href{https://doi.org/10.1088/1751-8113/46/21/214011}{\emph{J. Phys. A} {\bfseries 46} (2013) 214011} [\href{https://arxiv.org/abs/1112.1016}{{\ttfamily 1112.1016}}].

\bibitem{Gubser:1998bc}
S.S.~Gubser, I.R.~Klebanov and A.M.~Polyakov, \emph{{Gauge theory correlators from noncritical string theory}}, \href{https://doi.org/10.1016/S0370-2693(98)00377-3}{\emph{Phys. Lett. B} {\bfseries 428} (1998) 105} [\href{https://arxiv.org/abs/hep-th/9802109}{{\ttfamily hep-th/9802109}}].

\bibitem{Banks:1998dd}
T.~Banks, M.R.~Douglas, G.T.~Horowitz and E.J.~Martinec, \emph{{AdS dynamics from conformal field theory}},  \href{https://arxiv.org/abs/hep-th/9808016}{{\ttfamily hep-th/9808016}}.

\bibitem{Maldacena:2015waa}
J.~Maldacena, S.H.~Shenker and D.~Stanford, \emph{{A bound on chaos}}, \href{https://doi.org/10.1007/JHEP08(2016)106}{\emph{JHEP} {\bfseries 08} (2016) 106} [\href{https://arxiv.org/abs/1503.01409}{{\ttfamily 1503.01409}}].

\bibitem{Grozdanov:2017ajz}
S.~Grozdanov, K.~Schalm and V.~Scopelliti, \emph{{Black hole scrambling from hydrodynamics}}, \href{https://doi.org/10.1103/PhysRevLett.120.231601}{\emph{Phys. Rev. Lett.} {\bfseries 120} (2018) 231601} [\href{https://arxiv.org/abs/1710.00921}{{\ttfamily 1710.00921}}].

\bibitem{Blake:2017ris}
M.~Blake, H.~Lee and H.~Liu, \emph{{A quantum hydrodynamical description for scrambling and many-body chaos}}, \href{https://doi.org/10.1007/JHEP10(2018)127}{\emph{JHEP} {\bfseries 10} (2018) 127} [\href{https://arxiv.org/abs/1801.00010}{{\ttfamily 1801.00010}}].

\bibitem{Blake:2018leo}
M.~Blake, R.A.~Davison, S.~Grozdanov and H.~Liu, \emph{{Many-body chaos and energy dynamics in holography}}, \href{https://doi.org/10.1007/JHEP10(2018)035}{\emph{JHEP} {\bfseries 10} (2018) 035} [\href{https://arxiv.org/abs/1809.01169}{{\ttfamily 1809.01169}}].

\bibitem{Wang:2022mcq}
D.~Wang and Z.-Y.~Wang, \emph{{Pole Skipping in Holographic Theories with Bosonic Fields}}, \href{https://doi.org/10.1103/PhysRevLett.129.231603}{\emph{Phys. Rev. Lett.} {\bfseries 129} (2022) 231603} [\href{https://arxiv.org/abs/2208.01047}{{\ttfamily 2208.01047}}].

\bibitem{Perlmutter:2016pkf}
E.~Perlmutter, \emph{{Bounding the Space of Holographic CFTs with Chaos}}, \href{https://doi.org/10.1007/JHEP10(2016)069}{\emph{JHEP} {\bfseries 10} (2016) 069} [\href{https://arxiv.org/abs/1602.08272}{{\ttfamily 1602.08272}}].

\bibitem{Heemskerk:2009pn}
I.~Heemskerk, J.~Penedones, J.~Polchinski and J.~Sully, \emph{{Holography from Conformal Field Theory}}, \href{https://doi.org/10.1088/1126-6708/2009/10/079}{\emph{JHEP} {\bfseries 10} (2009) 079} [\href{https://arxiv.org/abs/0907.0151}{{\ttfamily 0907.0151}}].

\bibitem{Chua:2025vig}
W.Z.~Chua, T.~Hartman and W.W.~Weng, \emph{{Replica manifolds, pole skipping, and the butterfly effect}},  \href{https://arxiv.org/abs/2504.08139}{{\ttfamily 2504.08139}}.

\bibitem{Boulanger:2013zza}
N.~Boulanger, D.~Ponomarev, E.D.~Skvortsov and M.~Taronna, \emph{{On the uniqueness of higher-spin symmetries in AdS and CFT}}, \href{https://doi.org/10.1142/S0217751X13501625}{\emph{Int. J. Mod. Phys. A} {\bfseries 28} (2013) 1350162} [\href{https://arxiv.org/abs/1305.5180}{{\ttfamily 1305.5180}}].

\bibitem{Alba:2015upa}
V.~Alba and K.~Diab, \emph{{Constraining conformal field theories with a higher spin symmetry in $d > 3$ dimensions}}, \href{https://doi.org/10.1007/JHEP03(2016)044}{\emph{JHEP} {\bfseries 03} (2016) 044} [\href{https://arxiv.org/abs/1510.02535}{{\ttfamily 1510.02535}}].

\bibitem{Afkhami-Jeddi:2017idc}
N.~Afkhami-Jeddi, K.~Colville, T.~Hartman, A.~Maloney and E.~Perlmutter, \emph{{Constraints on higher spin CFT$_{2}$}}, \href{https://doi.org/10.1007/JHEP05(2018)092}{\emph{JHEP} {\bfseries 05} (2018) 092} [\href{https://arxiv.org/abs/1707.07717}{{\ttfamily 1707.07717}}].

\bibitem{Knysh:2024asf}
M.~Knysh, H.~Liu and N.~Pinzani-Fokeeva, \emph{{New horizon symmetries, hydrodynamics, and quantum chaos}}, \href{https://doi.org/10.1007/JHEP09(2024)162}{\emph{JHEP} {\bfseries 09} (2024) 162} [\href{https://arxiv.org/abs/2405.17559}{{\ttfamily 2405.17559}}].

\bibitem{Hartman:2016lgu}
T.~Hartman, S.~Kundu and A.~Tajdini, \emph{{Averaged Null Energy Condition from Causality}}, \href{https://doi.org/10.1007/JHEP07(2017)066}{\emph{JHEP} {\bfseries 07} (2017) 066} [\href{https://arxiv.org/abs/1610.05308}{{\ttfamily 1610.05308}}].

\bibitem{Kravchuk:2018htv}
P.~Kravchuk and D.~Simmons-Duffin, \emph{{Light-ray operators in conformal field theory}}, \href{https://doi.org/10.1007/JHEP11(2018)102}{\emph{JHEP} {\bfseries 11} (2018) 102} [\href{https://arxiv.org/abs/1805.00098}{{\ttfamily 1805.00098}}].

\bibitem{Meltzer:2018tnm}
D.~Meltzer, \emph{{Higher Spin ANEC and the Space of CFTs}}, \href{https://doi.org/10.1007/JHEP07(2019)001}{\emph{JHEP} {\bfseries 07} (2019) 001} [\href{https://arxiv.org/abs/1811.01913}{{\ttfamily 1811.01913}}].

\bibitem{Hayden:2007cs}
P.~Hayden and J.~Preskill, \emph{{Black holes as mirrors: Quantum information in random subsystems}}, \href{https://doi.org/10.1088/1126-6708/2007/09/120}{\emph{JHEP} {\bfseries 09} (2007) 120} [\href{https://arxiv.org/abs/0708.4025}{{\ttfamily 0708.4025}}].

\bibitem{Sekino:2008he}
Y.~Sekino and L.~Susskind, \emph{{Fast Scramblers}}, \href{https://doi.org/10.1088/1126-6708/2008/10/065}{\emph{JHEP} {\bfseries 10} (2008) 065} [\href{https://arxiv.org/abs/0808.2096}{{\ttfamily 0808.2096}}].

\bibitem{Klebanov:2002ja}
I.R.~Klebanov and A.M.~Polyakov, \emph{{AdS dual of the critical O(N) vector model}}, \href{https://doi.org/10.1016/S0370-2693(02)02980-5}{\emph{Phys. Lett. B} {\bfseries 550} (2002) 213} [\href{https://arxiv.org/abs/hep-th/0210114}{{\ttfamily hep-th/0210114}}].

\bibitem{Ceplak:2019ymw}
N.~Ceplak, K.~Ramdial and D.~Vegh, \emph{{Fermionic pole-skipping in holography}}, \href{https://doi.org/10.1007/JHEP07(2020)203}{\emph{JHEP} {\bfseries 07} (2020) 203} [\href{https://arxiv.org/abs/1910.02975}{{\ttfamily 1910.02975}}].

\bibitem{Ceplak:2021efc}
N.~Ceplak and D.~Vegh, \emph{{Pole-skipping and Rarita-Schwinger fields}}, \href{https://doi.org/10.1103/PhysRevD.103.106009}{\emph{Phys. Rev. D} {\bfseries 103} (2021) 106009} [\href{https://arxiv.org/abs/2101.01490}{{\ttfamily 2101.01490}}].

\bibitem{Ning:2023ggs}
S.~Ning, D.~Wang and Z.-Y.~Wang, \emph{{Pole skipping in holographic theories with gauge and fermionic fields}}, \href{https://doi.org/10.1007/JHEP12(2023)084}{\emph{JHEP} {\bfseries 12} (2023) 084} [\href{https://arxiv.org/abs/2308.08191}{{\ttfamily 2308.08191}}].

\bibitem{Gutperle:2011kf}
M.~Gutperle and P.~Kraus, \emph{{Higher Spin Black Holes}}, \href{https://doi.org/10.1007/JHEP05(2011)022}{\emph{JHEP} {\bfseries 05} (2011) 022} [\href{https://arxiv.org/abs/1103.4304}{{\ttfamily 1103.4304}}].

\bibitem{Ammon:2011nk}
M.~Ammon, M.~Gutperle, P.~Kraus and E.~Perlmutter, \emph{{Spacetime Geometry in Higher Spin Gravity}}, \href{https://doi.org/10.1007/JHEP10(2011)053}{\emph{JHEP} {\bfseries 10} (2011) 053} [\href{https://arxiv.org/abs/1106.4788}{{\ttfamily 1106.4788}}].

\bibitem{Lewkowycz:2013nqa}
A.~Lewkowycz and J.~Maldacena, \emph{{Generalized gravitational entropy}}, \href{https://doi.org/10.1007/JHEP08(2013)090}{\emph{JHEP} {\bfseries 08} (2013) 090} [\href{https://arxiv.org/abs/1304.4926}{{\ttfamily 1304.4926}}].

\bibitem{Dong:2023bax}
X.~Dong, G.N.~Remmen, D.~Wang, W.W.~Weng and C.-H.~Wu, \emph{{Holographic entanglement from the UV to the IR}}, \href{https://doi.org/10.1007/JHEP11(2023)207}{\emph{JHEP} {\bfseries 11} (2023) 207} [\href{https://arxiv.org/abs/2308.07952}{{\ttfamily 2308.07952}}].

\bibitem{Wall:2015raa}
A.C.~Wall, \emph{{A Second Law for Higher Curvature Gravity}}, \href{https://doi.org/10.1142/S0218271815440149}{\emph{Int. J. Mod. Phys. D} {\bfseries 24} (2015) 1544014} [\href{https://arxiv.org/abs/1504.08040}{{\ttfamily 1504.08040}}].

\bibitem{Yan:2024gbz}
Z.~Yan, \emph{{Gravitational focusing and horizon entropy for higher-spin fields}},  \href{https://arxiv.org/abs/2412.07107}{{\ttfamily 2412.07107}}.

\end{thebibliography}\endgroup

\end{document}